\documentstyle[aps,prb]{revtex}
\begin{document}
\draft
\title{
Metal-insulator transition of isotopically enriched 
neutron-transmutation-doped $^{70}$Ge:Ga 
in magnetic fields
}
\author{Michio Watanabe} 
\address{Department of Applied Physics and Physico-Informatics, Keio University, 
3-14-1 Hiyoshi, Kohoku-ku, Yokohama 223-8522, Japan}
\author{Kohei M. Itoh}
\address{Department of Applied Physics and Physico-Informatics, Keio University 
and PRESTO-JST, 3-14-1 Hiyoshi, Kohoku-ku, Yokohama 223-8522, Japan}
\author{Youiti Ootuka}
\address{Institute of Physics, University of Tsukuba, 
1-1-1 Tennodai, Tsukuba, Ibaraki 305-8571, Japan}
\author{Eugene E. Haller}
\address{Lawrence Berkeley National Laboratory and University of
California at Berkeley, Berkeley, California 94720}
\date{Received 17 June 1999}
\maketitle
\begin{abstract}
We have investigated the temperature dependence of the electrical conductivity
$\sigma(N,B,T)$ of nominally uncompensated, neutron-transmutation-doped
$^{70}$Ge:Ga samples in magnetic fields up to $B=8$~T
at low temperatures ($T=0.05-0.5$~K).
In our earlier studies at $B=0$, 
the critical exponent $\mu=0.5$ defined by 
$\sigma(N,0,0) \propto (N-N_c)^{\mu}$ has been determined 
for the same series of $^{70}$Ge:Ga samples with the doping concentration 
$N$ ranging from $1.861\!\times\!10^{17}\,{\rm cm}^{-3}$ to 
$2.434\!\times\!10^{17}\,{\rm cm}^{-3}$.  
In magnetic fields, the motion of carriers loses 
time-reversal symmetry, the universality class may change 
and with it the value of $\mu$.  
In this work, we show that magnetic fields indeed affect 
the value of $\mu$ ($\mu$ changes from 0.5 at $B=0$ to 1.1 at $B\geq4$~T).  
The same exponent $\mu'=1.1$ is also found in the 
magnetic-field-induced MIT for three different $^{70}$Ge:Ga samples, i.e.,
$\sigma(N,B,0) \propto [B_c(N)-B]^{\mu'}$ where $B_c(N)$ is 
the concentration-dependent critical magnetic induction.
We show that $\sigma(N,B,0)$ obeys a simple scaling rule 
on the $(N,B)$~plane.  Based on this finding, we derive from a simple 
mathematical argument that $\mu=\mu'$ as has been observed in our experiment.  
\end{abstract}

\section{Introduction}
\label{sec:intro}
Semiconductors with a random distribution of doping impurities 
have been studied extensively over the past three decades 
in order to probe the nature of the metal-insulator transition (MIT) 
in disordered electronic systems.~\cite{Lee85,Bel94}  
The value of the critical exponent $\mu$ of the conductivity for the 
metallic side of the transition, however, still remains controversial.  
The exponent $\mu$ is defined by 
\begin{equation}
\label{eq:critM}
\sigma (0) = \sigma^* (N/N_c - 1)^\mu
\end{equation}
in the critical regime of the MIT $(0<N/N_c-1\ll1)$.  
Here, $\sigma(0)$ is the zero-temperature 
conductivity, $\sigma^*$ is a prefactor, 
$N$ is the impurity concentration, and $N_c$ is the critical 
concentration for the MIT. 
An exponent of $\mu\approx0.5$ has been found 
in a number of nominally uncompensated semiconductors 
[Si:P,~\cite{Ros83} Si:As,~\cite{Sha89} Si:Sb,~\cite{Lon84} 
Ge:As,~\cite{Ion91} and $^{70}$Ge:Ga~(Refs.~\ref{Ito96} and 
\ref{Wat98})].  
This value is considerably smaller than the results 
of numerical calculations ($\mu=1.2\sim1.6$; 
e.g., Refs.~\ref{Mac81} and \ref{Sle97}) 
for the Anderson transition purely driven by disorder.  
Therefore, electron-electron interaction, 
which is undoubtedly present in doped semiconductors, 
must be relevant to the nature of the MIT, 
at least when impurity compensation is absent.  
A conclusion that has been reached over the years is that 
one has to deal simultaneously with disorder 
and electron-electron interaction in order 
to understand the MIT in doped semiconductors.  

According to theories~\cite{Bel94} on the MIT 
which take into account both disorder and 
electron-electron interaction, the critical exponent 
$\mu$ does not depend on the details of the system, but depends 
only on the universality class to which the system belongs.  
Moreover, there is an inequality $\nu\geq2/3$ 
for the critical exponent $\nu$ of the correlation length.~\cite{Cha86}  
The inequality is expected to apply generally to disordered systems, 
irrespective of the presence of electron-electron interaction.~\cite{Paz97}  
Hence, if one assumes the Wegner relation~\cite{Weg76} $\mu=\nu$, 
which is derived for systems {\em without} electron-electron 
interaction, $\mu\approx0.5$ violates the inequality.  
This discrepancy has been known as the conductivity critical 
exponent puzzle.  Kirkpatrick and Belitz~\cite{Kir93} 
have claimed that there are logarithmic corrections to scaling 
in universality classes with time-reversal symmetry, 
i.e., when the external magnetic field is zero, and that  
$\mu\approx0.5$, found at $B=0$, should be interpreted as an ``effective" 
exponent which is different from a ``real" exponent satisfying $\mu\geq2/3$.  
Therefore, comparison of $\mu$ with and without the time-reversal symmetry, 
i.e., with and without external magnetic fields becomes important.  
Experimentally, $\mu\approx 1$ has been found for magnetic inductions $B$ 
on the order of one tesla for nominally uncompensated semiconductors: 
Ge:Sb,~\cite{Oot88,Ros89} Si:B,~\cite{Dai92} and Si:P~(Ref.~\ref{Dai93}).  
Since these systems result in different values of $\mu$ ranging from 0.5 to 1.0 
at $B=0$, the applied magnetic field changes the value of $\mu$ for certain 
systems (Si:B and Si:P), while it does not~\cite{Ros89} or does only change 
little~\cite{Oot88} for the other (Ge:Sb). In this work, we aim to achieve 
a complete understanding of the effect of magnetic fields on the MIT 
in uncompensated semiconductors by studying the critical behavior 
of the zero-temperature conductivity 
as a function of both $N$ (doping-induced MIT) and $B$ 
(magnetic-field-induced MIT) in magnetic fields up to 8~T 
for $^{70}$Ge:Ga system.  
To our knowledge, the MIT in Si or Ge has not been analyzed 
as a function of $B$.  
Concerning the critical point, 
\begin{equation}
N_c(B)-N_c(0) \propto B^{\beta} 
\end{equation}
with $\beta=0.5$ 
was obtained for Ge:Sb,~\cite{Ros89} while $\beta=1.7\pm0.4$ 
for Si:B.~\cite{Bog97}  The exponent $\beta$ characterizes the phase 
(metal or insulator) diagram on the $(N,B)$~plane, and provides information 
on the nature of the MIT in magnetic fields.  The above experimental results, 
however, imply that $\beta$ could be completely different even though 
$\mu$ in magnetic fields is the same.  Hence, a determination of $\beta$ 
for various systems is important in order to probe the effect of 
magnetic fields.  

In our earlier studies,~\cite{Ito96,Wat98} we obtained $\mu=0.5$ 
at $B=0$ for $^{70}$Ge:Ga.  This result was obtained from precisely doped 
samples with a perfectly random distribution of impurities; our $^{70}$Ge:Ga 
samples were prepared by neutron-transmutation doping (NTD), 
in which an ideally random distribution of dopants is inherently 
guaranteed down to the atomic level.~\cite{Hal84,Par88,Hal90,Ito93}  
For the case of melt- (or metallurgically) doped samples 
that have been employed in most of the previous 
studies,~\cite{Ros83,Sha89,Lon84,Oot88,Ros89,Dai92,Dai93,Bog97} 
the spatial fluctuation of $N$ due to dopant striations 
and segregation can easily be on the order of 1\% or more across 
a typical sample for the four-point resistance measurement 
(length of $\sim$5~mm or larger),~\cite{Shi88}  
and hence, it will not be meaningful to discuss 
physical properties in the critical regime 
(e.g., $|N/N_c-1| < 0.01$), unless one evaluates  
the macroscopic inhomogeneity in the samples and 
its influence on the results.
A homogeneous distribution of impurities is important 
also for experiments in magnetic fields.  Using the same series 
of $^{70}$Ge:Ga samples that was employed in our previous 
study,~\cite{Wat98} we show here that the critical exponent of 
the conductivity is 1.1 in magnetic fields for both the doping-induced MIT 
and the magnetic-field-induced MIT.  
The phase diagram on the $(N,B)$~plane is successfully constructed, 
and $\beta=2.5$ is obtained for $^{70}$Ge:Ga.

\section{Experiment}
\label{sec:ex}
All of the $^{70}$Ge:Ga samples were prepared 
by neutron-transmutation doping (NTD) 
of isotopically enriched $^{70}$Ge single crystals.  
We use the NTD process since it is known to produce 
the most homogeneous, perfectly random dopant distribution 
down to the atomic level.~\cite{Hal84,Par88,Hal90,Ito93} 
The concentration $N$ of Ga acceptors is determined from 
the time of irradiation with thermal neutron.  The concentration $N$ 
is proportional to the irradiation time as long as the same irradiation 
site and the same power of a nuclear reactor are employed.  
Details of the sample preparation and characterization are described 
elsewhere.~\cite{Wat98}  In this work 12 samples that are metallic 
in zero magnetic field are studied.  (See Table~\ref{tab:samp}.)  
The conductivity of the samples in zero magnetic field 
has been reported in Refs.~\ref{Ito96} and \ref{Wat98}.  

We determined the electrical conductivity of the samples 
at low temperatures between 0.05~K and 0.5~K  
using a $^{3}$He-$^{4}$He dilution refrigerator.  
Magnetic fields up to 8~T were applied in the direction perpendicular 
to the current flow by means of a superconducting solenoid. 

\section{Results}
\label{sec:results}
\subsection{Temperature dependence of conductivity}
Figure~\ref{fig:B2} shows the temperature dependence of the conductivity 
of Sample~B2 for several values of the magnetic induction $B$.  
Application of the magnetic field decreases the conductivity 
and eventually drives the sample into the insulating phase.  
This property can be understood in terms of the shrinkage 
of the wave function due to the magnetic field.  

The temperature variation of the conductivity $\sigma(N,B,T)$ 
of a disordered metal at low temperatures is governed mainly 
by electron-electron interaction,~\cite{Lee85} and can be 
written in zero magnetic field as 
\begin{equation}
\label{eq:1/2}
\sigma (N,0,T) = \sigma (N,0,0) + m(N,0)\,T^{1/2}\,.
\end{equation}
When $g \mu_B B \gg k_B T$, i.e., in strong magnetic fields at low 
temperatures, the conductivity shows another $T^{1/2}$ dependence 
\begin{equation}
\label{eq:eeinB1}
\sigma(N,B,T) = \sigma(N,B,0) + m_B(N,B)\,T^{1/2}\,.
\end{equation}
Here, one should note that these equations are valid only 
in the limits of $[\sigma (N,0,T) - \sigma (N,0,0)] \ll \sigma(N,0,0)$ 
or $[\sigma(N,B,T) - \sigma(N,B,0)] \ll \sigma(N,B,0)$. 
It is for this reason that we have observed a $T^{1/3}$ dependence 
rather than the $T^{1/2}$ dependence at $B=0$ in $^{70}$Ge:Ga 
as the critical point [$\sigma(N,0,0)=0$] is approached 
from the metallic side.~\cite{Wat98}  
However, Fig.~\ref{fig:B2} shows that the $T^{1/2}$ dependence 
holds when $B\neq0$ even around the critical point. 
Hence, we use Eq.~(\ref{eq:eeinB1}) to evaluate the zero-temperature 
conductivity $\sigma(N,B,0)$ in magnetic fields.  

According to an interaction theory for a disordered metal,~\cite{Lee85} 
$m$ and $m_B$ are given by 
\begin{equation}
\label{eq:m}
m \,=\, \frac{\;e^2}{\hbar}\frac{1}{\;4\pi^2\;}\frac{1.3}{\;\sqrt{2}\;}
\left(\frac{4}{\;3\;}-\frac{3}{\;2\;}\tilde{F} \right) 
\sqrt{\frac{k_B}{\;\hbar D\;}}
\end{equation}
and 
\begin{equation}
\label{eq:mB}
m_B \,=\, \frac{\;e^2}{\hbar}\frac{1}{\;4\pi^2\;}\frac{1.3}{\;\sqrt{2}\;}
\left(\frac{4}{\;3\;}-\frac{1}{\;2\;}\tilde{F} \right) 
\sqrt{\frac{k_B}{\;\hbar D\;}}\;,
\end{equation}
respectively,  
where $D$ is the diffusion constant and $\tilde{F}$ is 
a dimensionless parameter characterizing the Hartree interaction.  
Since $m_B$ is independent of $B$, the conductivity for various values 
of $B$ plotted against $T^{1/2}$ should appear as a group of parallel lines.  
This is approximately the case as seen in Fig.~\ref{fig:B2} at low 
temperatures (e.g., $T<0.25$~K).  Values of $m$ (for $B=0$) and $m_B$ 
at $B=4$~T differ from each other considerably 
for all the samples as listed in Table~\ref{tab:samp}.  
This implies that $\tilde{F}$ is of the order of unity 
according to Eqs.~(\ref{eq:m}) and (\ref{eq:mB}).  
In order to support this finding, we shall estimate $\tilde{F}$ 
within the context of the Thomas-Fermi approximation.  

The parameter $\tilde{F}$ is related to the average $F$ of 
the screened Coulomb interaction on the Fermi surface as 
\begin{equation}
\tilde{F} = -\frac{\;32\;}{3}
\left[\frac{\;1 + 3F/4 - (1+F/2)^{3/2}\;}{F}\right]\;.
\end{equation}
The Thomas-Fermi approximation gives 
\begin{equation}
F = \frac{\,\ln (1+x)\,}{x}\;,
\end{equation}
where 
\begin{equation}
x = (2k_F/\kappa)^2,
\end{equation}
with the Fermi wave vector
\begin{equation}
k_F = (3\pi^2N)^{1/3}, 
\end{equation}
and the screening wave vector in SI units
\begin{equation}
\kappa = \sqrt{3e^2Nm^*/(\epsilon \epsilon_0 \hbar^2 k_F^2)}\;.
\end{equation}
For Ge, the relative dielectric constant $\epsilon$ is 15.8 
and the effective mass $m^*$ of a heavy hole is $0.34m_e$, 
where $m_e$ is the electron rest mass.~\cite{Kit86} 
Hence $x=1.1\tilde{N}^{1/3}$, where $\tilde{N}$ 
is in units of 10$^{17}$ cm$^{-3}$. 
Thus, in the concentration range covered by the samples, 
the Thomas-Fermi approximation gives $0.48<\tilde{F}<0.55$, 
which is consistent with the experimental finding that $\tilde{F}$ 
is of the order of unity.  

\subsection{Doping-induced metal-insulator transition}
The zero-temperature conductivity $\sigma(N,B,0)$ of the $^{70}$Ge:Ga 
samples in various magnetic fields obtained by extrapolation of 
$\sigma(N,B,T)$ to $T=0$ based on Eq.~(\ref{eq:eeinB1}) is shown 
in Fig.~\ref{fig:nMIT}.
Here, $\sigma(N,B,0)$ is plotted as a function of the normalized concentration: 
\begin{equation}
\label{eq:n}
n \equiv [\sigma(N,0,0)/\sigma^*(0)]^{2.0}.  
\end{equation}
Since the relation between $N$ and $\sigma(N,0,0)$ was 
established for $^{70}$Ge:Ga in Ref.~\ref{Wat98} as
$\sigma(N,0,0) = \sigma^*(0)[N/N_c(0)-1]^{0.50}$ 
where $N_c(0)=1.861\!\times\!10^{17}\,{\rm cm}^{-3}$, 
$n$ is equivalent to $N/N_c(0)-1$.  
Henceforth, we will use $n$ instead of $N$ because employing $n$ 
reduces the scattering of the data caused by several experimental 
uncertainties, and it further helps us concentrate on observing how 
$\sigma(N,B,0)$ varies as $B$ is increased.  
Similar evaluations of the concentration have been used by various groups.  
In their approach, the ratio of the resistance at 4.2~K to that at 300~K is 
used to determine the concentration.~\cite{Dai92}  
The dashed curve in Fig.~\ref{fig:nMIT} is for $B=0$, which 
merely expresses Eq.~(\ref{eq:n}), 
and the solid curves represent fits of 
\begin{equation}
\label{eq:nMIT}
\sigma(N,B,0) = \sigma_0(B)[n/n_c(B)-1]^{\mu(B)}.  
\end{equation}
The exponent $\mu(B)$ increases from 0.5 with increasing $B$ 
and reaches a value close to unity at $B\geq4$~T.  For example, 
$\mu=1.03\pm0.03$ at $B=4$~T and $\mu=1.09\pm0.05$ at $B=5$~T.  
When $B\geq6$~T, three-parameter [$\sigma_0(B)$, 
$n_c(B)$, and $\mu(B)$] fits no longer give reasonable results  
because the number of samples available for the fit 
decreases with increasing $B$.  Hence, we give the solid curves 
for $B\geq6$~T assuming $\mu(B)=1.15$.  

\subsection{Magnetic-field-induced metal-insulator transition}
\label{sec:R-BMIT}
We show $\sigma(N,B,0)$ as a function of $B$ 
in Fig.~\ref{fig:BMIT} for three different samples.  
When the magnetic field is weak, i.e., the correction 
$\Delta\sigma_B(N,B,0)\equiv\sigma(N,B,0)-\sigma(N,0,0)$ 
due to $B$ is small compared with $\sigma(N,0,0)$, the field 
dependence of $\Delta\sigma_B(N,B,0)$ looks consistent with 
the prediction by the interaction theory,~\cite{Lee85} 
\begin{equation}
\label{eq:sB0}
\Delta\sigma_B(N,B,0)\;=\;-\frac{\;e^2}{\hbar}
\frac{\tilde{F}}{\;4\pi^2\;}
\sqrt{\frac{\;g\mu_BB\;}{2\hbar D}}\;\propto\sqrt{B}\,.
\end{equation} 
In larger magnetic fields, $\sigma(N,B,0)$ deviates 
from Eq.~(\ref{eq:sB0}) and eventually vanishes 
at some magnetic induction $B_c$.  
For the samples in Fig.~\ref{fig:BMIT}, we tuned the magnetic 
induction to the MIT in a resolution of 0.1~T.  
We fit an equation similar to Eq.~(\ref{eq:nMIT}),
\begin{equation}
\label{eq:BMIT}
\sigma(N,B,0) 
= \sigma_{0}'(n)[1-B/B_c(n)]^{\mu'(n)},  
\end{equation}
to the data close to the critical point.  
As a result we obtain $\mu'=1.1\pm0.1$ for all of the three samples.  
The value of $\mu'$ depends on the choice of the magnetic-field range 
to be used for the fitting, and this fact leads to the error of $\pm0.1$ 
in the determination of $\mu'$.  

\subsection{Phase diagram in magnetic fields}
From the critical points $n_c(B)$ and $B_c(n)$, the phase diagram at $T=0$
is constructed on the $(N,B)$~plane as shown in Fig.~\ref{fig:nc}. 
Here, $n_c(B)$ for $B\geq6$~T shown by triangles 
are obtained by assuming $\mu=1.15$.  The vertical solid lines associated 
with the triangles represent the range of values over which $n_c(B)$ have 
to exist, i.e., between the highest $n$ in the insulating phase 
and the lowest $n$ in the metallic phase.  Solid diamonds represent 
$B_c$ for the three samples in which we have studied 
the magnetic-field-induced MIT in the preceding subsection.  
Estimations of $B_c$ for the other samples are also shown 
by open boxes with error bars.  

The boundary between metallic phase and insulating phase is expressed 
by a power-law relation: 
\begin{equation}
\label{eq:boundary}
n = C\,B^\beta .
\end{equation}
From the eight data points denoted by the solid symbols, 
we obtain $C=(1.33 \pm 0.17)\times 10^{-3}$~T$^{-\beta}$ 
and $\beta=2.45\pm0.09$ as shown by the dotted curve.  

\section{Discussion}
\subsection{Scaling of zero-temperature conductivity 
         in magnetic fields} 
Now we shall consider the relationship between the two critical 
exponents: $\mu$ for the doping-induced MIT and $\mu'$ for the 
magnetic-field-induced MIT.  
Suppose that a sample with normalized concentration $n$ 
has a zero-temperature conductivity $\sigma$ at $B\neq0$ 
and that $[n/n_c(B)-1]\ll 1$ or $[1-B/B_c(n)]\ll 1$.  
From Eqs.~(\ref{eq:nMIT}) and (\ref{eq:BMIT}), 
we have two expressions for $\sigma$:  
\begin{equation}
\label{eq:n1MIT}
\sigma = \sigma_0 \,(n/n_c-1)^\mu
\end{equation}
and
\begin{equation}
\label{eq:B1MIT}
\sigma = \sigma_0'\,(1-B/B_c)^{\mu'}.
\end{equation}
On the other hand, we have from Eq.~(\ref{eq:boundary}) 
\begin{equation}
\label{eq:approx1}
n/n_c = (B/B_c)^{-\beta} = [1-(1-B/B_c)]^{-\beta}
\approx1+\beta(1-B/B_c)
\end{equation}
in the limit of $(1-B/B_c)\ll1 $. 
This equation can be rewritten as 
\begin{equation}
\label{eq:approx2}
 (n/n_c-1)/\beta \,\approx\, (1-B/B_c). 
\end{equation}
Using Eqs.~(\ref{eq:n1MIT}), (\ref{eq:B1MIT}), 
and (\ref{eq:approx2}), we obtain 
\begin{equation}
\label{eq:always}
\sigma_0'(1-B/B_c)^{\mu'} \approx \beta^\mu\sigma_0 (1-B/B_c)^\mu. 
\end{equation}
Since Eq.~(\ref{eq:always}) has to hold for arbitrary $B$, 
the following relations 
\begin{equation}
\label{eq:s0's0}
\sigma_0'=\beta^\mu\sigma_0
\end{equation}
and 
\begin{equation}
\label{eq:mu'mu}
\mu'=\mu
\end{equation}
are derived. 

In Fig.~\ref{fig:scaling}, we see how well Eq.~(\ref{eq:mu'mu}) holds for 
the present system.  In Sec.~\ref{sec:R-BMIT}, we have already shown 
that $\mu'= 1.1 \pm 0.1$ is practically independent of $n$.
Concerning the exponent $\mu$, however, its dependence on $B$ has not 
been ruled out completely even for the highest $B$ we used 
in the experiments.  This is mainly because the number of available data points 
at large $B$ is not sufficient for a precise determination of $\mu$.  
In Fig.~\ref{fig:scaling}, 
the results of the doping-induced MIT for $B\geq4$~T 
(solid symbols) and the magnetic-field-induced MIT 
for three different samples (open symbols) are plotted.  
Here, we plot $\sigma(N,B,0)/[\beta^\mu\,\sigma_0(B)]$ vs 
$[n/n_c(B)-1]/\beta$ with $\beta=2.5$ and $\mu=1.1$ 
for the doping-induced MIT, and $\sigma(N,B,0)/\sigma_0(B)'$ 
vs $[1-B/B_c(n)]$ for the magnetic-field-induced MIT.  
Figure~\ref{fig:scaling} clearly shows that the data points 
align exceptionally well along a single line describing a single exponent 
$\mu=\mu'=1.1$.  

We saw in Fig.~\ref{fig:nMIT} that $\mu$ apparently takes 
smaller values in $B\leq3$~T, which seemingly contradicts 
the above consideration. We can understand this as follows.  
We find that the critical exponent $\mu$ in zero magnetic field 
is 0.5 which is different from the values of $\mu$ in magnetic fields.
Hence, one should note whether the system under consideration 
belongs to the ``magnetic-field regime" or not.
In systems where the MIT occurs, there are several characteristic 
length scales: the correlation length, the thermal diffusion length, 
the inelastic scattering length, the spin scattering length,
the spin-orbit scattering length, etc.
As for the magnetic field, it is characterized by the magnetic length 
$\lambda\equiv\sqrt{\hbar/eB}$.  When $\lambda$ is smaller than 
the other length scales, the system is in the ``magnetic-field regime."
As the correlation length $\xi$ diverges at the MIT, $\lambda < \xi$ holds 
near the critical point, no matter how weak the magnetic field is. 
When the field is not sufficiently large, the ``magnetic-field regime" 
where we assume $\mu = 1.1$ to hold, is restricted to a narrow region of 
concentration.  Outside the region, the system crosses over to 
the ``zero-field regime" where $\mu = 0.5$ is expected.  
This is what is seen in Fig.~\ref{fig:nMIT}.  

The constant critical exponent in $B\neq0$ yields a scaling of the form
\begin{equation}
\label{eq:scal}
\sigma(N,B,0)=\tilde{\sigma}(n,B)\:f(n/B^\beta)\,,
\end{equation}
where $\tilde{\sigma}(n,B)$ is a prefactor which is irrelevant to
the transition.  
The values of the prefactor are listed in Table~\ref{tab:pre}.  
Here, we list $\sigma^*$  in Eq.~(\ref{eq:critM}) instead of $\sigma_0$;  
$\sigma^*$~in~$B\neq0$ is calculated from $\sigma_0$ as
\begin{equation}
\sigma^*= (1+n_c^{-1})^\mu\;\sigma_0
\end{equation}
because the relation between $n/n_c-1$ 
and $N/N_c-1$ is given by 
\begin{equation}
(n/n_c - 1) = (1+n_c^{-1})(N/N_c - 1)\,.
\end{equation}
The values of $\sigma^*$ for zero field and for other doped semiconductors
are also given in Table~\ref{tab:pre}.
The values are normalized to Mott's minimum metallic 
conductivity defined by 
\begin{equation}
\label{eq:s_min}
\sigma_{\rm min} \equiv C_{\rm M}(e^2/\hbar)N_c^{1/3}, 
\end{equation}
where we assume $C_{\rm M} = 0.05$ 
as Rosenbaum {\em et al.}~\cite{Ros83} did.  
The prefactor $\sigma^*$ can be also defined 
for the magnetic-field-induced MIT.  We proposed to define 
it based on Eqs.~(\ref{eq:s0's0}) and (\ref{eq:mu'mu}) as 
\begin{equation}
\sigma^*\equiv [(1+n^{-1})/\beta]^{\mu'}\;\sigma_0'\;.
\end{equation}
Using this definition, we calculated $\sigma^*$ for the three 
$^{70}$Ge:Ga samples in which we studied the magnetic-field-induced 
MIT.  The ratios $\sigma^*/\sigma_{\rm min}$ are 10, 8, and 6 for 
Samples~A3, B2, and B4, respectively.  It is reasonable 
that the ratios $\sigma^*/\sigma_{\rm min}$ for the three samples 
and those in Table~\ref{tab:pre} are of the same order of magnitude.  
Note that Mott's minimum metallic conductivity $\sigma_{\rm min}$ 
depends on both $B$ and the system through the critical concentration 
$N_c$.  [See Eq.~(\ref{eq:s_min}).]  

A similar scaling form was studied theoretically by 
Khmel'nitskii and Larkin.~\cite{Khm81}  They considered 
a noninteracting electron system starting from 
\begin{equation}
\label{eq:KL}
\sigma(N,B,0)\approx \frac{\,e^2}{\hbar\,\xi}\:f(B^\alpha\,\xi), 
\end{equation}
where $\xi$ is the correlation length.  They claimed that the argument of 
the function $f$ should be a power of the magnetic flux through a region 
with dimension $\xi$.  This means 
\begin{equation}
\label{eq:KL2}
\sigma(N,B,0)\approx \frac{\,e^2}{\hbar\,\xi}\:f(\xi/\lambda), 
\end{equation}
where $\lambda\equiv\sqrt{\hbar/eB}$ is the magnetic length,  
and hence, $\alpha=1/2$.  
In order to discuss the shift of the MIT due to the magnetic field,
they rewrote Eq.~(\ref{eq:KL2}) as 
\begin{equation}
\label{eq:KL5}
\sigma(N,B,0)\approx \frac{\,e^2}{\hbar\,\lambda}\:
                     \phi (t \, \lambda^{1/\nu}),
\end{equation}
based on the relation in zero magnetic field
\begin{equation}
\label{eq:xi}
\xi\propto t^{-\nu}.
\end{equation}
Here, $t$ is a measure of distance from the critical point 
in zero field, e.g., 
\begin{equation}
t\equiv[N/N_c(0)-1].
\end{equation}
The zero point of the function $\phi$ gives the MIT, and 
the shift of the critical point for the MIT equals  
\begin{equation}
N_c(B)-N_c(0) \propto B^{1/2\nu}.
\end{equation}
Thus, $\beta = 1/(2\nu)$ results.  
Rosenbaum, Field, and Bhatt~\cite{Ros89} 
reported $\beta=0.5$ and $\mu=1$ in Ge:Sb, 
which satisfies this relation, when one assumes 
the Wegner relation~\cite{Weg76} $\mu=\nu$.  
In the present system, however, 
this relation does not hold, as long as we assume the Wegner relation.  
Experimentally, we find $\beta=2.5$, while $1/(2\nu)=1/(2\mu)=1$
for $^{70}$Ge:Ga at $B=0$.  
The relation does not hold in Si:B, either ($\beta=1.7$ while 
$\mu=0.65$ at $B=0$).~\cite{Bog97}  

\subsection{Critical exponents}
\label{sec:D-mu}
Finally, we shall discuss the possible origin for the crossover of $\mu=0.5$ 
at $B=0$ to $\mu=1.1$ at $B\neq0$.  
According to theories~\cite{Bel94} for the MIT dealing 
with both disorder and electron-electron interaction, 
systems can be categorized into four universality classes
by the symmetry they have as listed in Table~\ref{tab:UC} 
and the value of the critical exponent depends only 
on the universality class. 
For classes~MF, MI, and SO, the nonlinear sigma model receives 
some restrictions from the symmetry breaker and the critical 
exponent $\nu$ for the correlation length is calculated for 
$d=2+\varepsilon$ dimensions as 
\begin{equation}
\label{eq:nu(2+e)}
\nu=\frac{1}{\;\varepsilon\;}\,[1+O(\varepsilon)]\;. 
\end{equation}
Assuming the Wegner relation~\cite{Weg76} 
\begin{equation}
\label{eq:Wegner}
\mu=\nu(d-2)\,,
\end{equation}
Eq.~(\ref{eq:nu(2+e)}) yields 
\begin{equation}
\mu=1+O(\varepsilon)\,, 
\end{equation}
which means $\mu\approx1$ when $\varepsilon\ll 1$.
For $d=3$ $(\varepsilon=1)$, however, the theoretical result tells 
us very little about the value of $\mu$. 
A calculation for class~G is difficult because there is no 
restriction for the nonlinear sigma model.  The value in 
Table~\ref{tab:UC} for $d=3$ is merely an {\em approximate} one. 
We believe that $``\nu\cong0.75"$ should be treated as 
less accurate than $``\nu=1+O(1)"$ for the other classes.  
So, we conclude that values of $\mu$ can only be determined at this time 
by experimental measurements.  

Ruling out an ambiguity due to 
an inhomogeneous distribution of impurities, 
we have established $\mu=0.5$ at $B=0$ 
and $\mu=1.1$ at $B\neq0$ for $^{70}$Ge:Ga.  
Since $^{70}$Ge:Ga is a $p$-type semiconductor, it is most 
likely categorized as class~SO.  Electrical transport properties of 
$p$-type semiconductors are governed by holes at the top of 
the valence band, where spin-orbit coupling partially removes 
the degeneracy and shifts the split-off band down in energy 
by 0.29~eV in Ge and 0.044~eV in Si.~\cite{Kit86}  
External magnetic fields change the universality class to which 
a system belongs by breaking the time-reversal symmetry.  
According to Ref.~\ref{Bel94}, a system in class~SO at $B=0$ 
belongs to class~MI at $B\neq0$ even if it contains no magnetic
impurities.  The change of $\mu=0.5$ to 
$\mu=1.1$ observed in $^{70}$Ge:Ga due to the application of a magnetic 
field should be understood in such a context. 
A similar phenomenon was also found in Si:B.~\cite{Dai92} 
(See Table~\ref{tab:pre}.) 

\section{Conclusion}
\label{sec-conc}
We have measured the electrical conductivity of NTD $^{70}$Ge:Ga samples 
in magnetic fields up to $B=8$~T in order to study 
the doping-induced MIT (in magnetic fields) 
and the magnetic-field-induced MIT.  
For both of the MIT, the critical exponent of the conductivity is 1.1, 
which is different from the value 0.5 at $B=0$.  
The change of the critical exponent caused by the applied magnetic fields 
supports a picture in which $\mu$ varies depending on the universality class 
to which the system belongs.  
The phase diagram has been determined in magnetic 
fields for the $^{70}$Ge:Ga system.  

\section*{Acknowledgments}
We are thankful to T. Ohtsuki for valuable discussions, J. W. Farmer 
for the neutron irradiation, and V. I. Ozhogin for the supply of the Ge 
isotope.  All the low-temperature measurements were carried out 
at the Cryogenic Center, the University of Tokyo.  
M. W. would like to thank Japan Society for the Promotion of Science 
(JSPS) for financial support.  
The work at Keio was supported in part by a Grant-in-Aid for 
Scientific Research from the Ministry of Education, Science, Sports, 
and Culture, Japan.  The work at Berkeley was supported in part 
by the Director, Office of Energy Research, Office of 
Basic Energy Science, Materials Sciences Division of the U. S. 
Department of Energy under Contract No.~DE-AC03-76SF00098 and in part 
by U. S. NSF Grant No.~DMR-97 32707.  


%
%
\begin{figure}
\caption{Conductivity of Sample~B2 as a function of $T^{1/2}$ 
at several magnetic fields.  The values of the magnetic induction 
from top to bottom in units of tesla are 0.0, 1.0, 2.0, 3.0, 4.0, 
4.7, 5.0, 5.3, 5.6, 6.0, 7.0, and 8.0, respectively.}  
\label{fig:B2}
\end{figure}
\begin{figure}
\caption{Zero-temperature conductivity $\sigma(N,B,0)$ 
         vs normalized concentration 
         $n \equiv [\sigma(N,0,0)/\sigma^*(0)]^{2.0} = N/N_c(0)-1$, 
         where $\sigma(N,0,0)$ is the zero-temperature conductivity 
         and $\sigma^*(0)$ is the prefactor both at $B=0$. 
         From top to bottom the magnetic induction increases from 1~T to 8~T 
         in steps of 1~T.  The dashed curve at the top is for 
         $B=0$.  The solid curves represent 
         fits of $\sigma(N,B,0) \propto 
         [n/n_c(B)-1]^{\mu(B)}$.  
         For $B\geq6$~T, we assume $\mu=1.15$.  }
\label{fig:nMIT}
\end{figure}
\begin{figure}
\caption{Zero-temperature conductivity $\sigma(N,B,0)$ 
         of Samples~A3, B2, and B4 vs magnetic induction $B$.}
\label{fig:BMIT}
\end{figure}
\begin{figure}
\caption{Phase diagram of $^{70}$Ge:Ga at $T=0$.  
The solid circles and the open triangles represent the critical 
concentrations $n_c$, and the solid diamonds and the open boxes 
the critical magnetic induction $B_c$.}
\label{fig:nc}
\end{figure}
\begin{figure}
\caption{Normalized zero-temperature conductivity $\sigma(N,B,0)
/\sigma_0'(n)$ and $\sigma(N,B,0)/[\beta^\mu\,\sigma_0(B)]$ 
as functions of $[1-B/B_c(n)]$ and $[n/n_c(B)-1]
/\beta$, respectively, where $\beta=2.5$ and $\mu=1.1$. 
The solid line denotes a power-law behavior with the exponent of 1.1.  
The open and solid symbols represent the results of 
the magnetic-field-induced metal-insulator transition (MIT) 
in the range $(1-B/B_c)<0.5$ for three different samples 
(A3, B2, and B4) and the doping-induced MIT 
in constant magnetic fields (4, 5, 6, 7, and 8~T), respectively.}  
\label{fig:scaling}
\end{figure}

\pagebreak
%
%
%
%
%
\begin{table}
\caption{
List of $^{70}$Ge:Ga samples employed in this study.  
$N$ is the concentration of gallium determined 
from the irradiation time and the flux of thermal neutron 
[the critical concentration for the metal-insulator transition (MIT) 
is $1.860\!\times\!10^{17}\,{\rm cm}^{-3}$
(Ref.~\protect{\ref{Wat98}})]; 
$\sigma(0,0)$ is the zero-temperature conductivity 
at $B=0$; $n$ is the normalized concentration 
defined by Eq.~(\protect{\ref{eq:n}}); 
$m$ is the temperature coefficient of the conductivity 
at $B=0$ given by Eq.~(\protect{\ref{eq:1/2}});  
$B_c$ is the critical magnetic induction for the magnetic-field-induced MIT; 
$\sigma(4\,{\rm T},0)$ is the zero-temperature conductivity 
at $B=4$~T; $m_B(4\,{\rm T})$ is the temperature coefficient 
at $B=4$~T which is similar to $m$. 
}
\label{tab:samp}
\begin{tabular}{cccccccc}
& $N$ & $\sigma(0,0)$ & & $m$ & $B_c$ &
$\sigma(4\,{\rm T},0)$ & $m_B(4\,{\rm T})$ \\ 
Sample & (10$^{17}$~cm$^{-3}$) & (S/cm) & $n$ & 
(S$\,$cm$^{-1}\,$K$^{-1/2}$)  & (T) & 
(S/cm) & (S$\,$cm$^{-1}\,$K$^{-1/2}$) \\ \hline 
\makebox[0pt][r]{A}\makebox[0pt][l]{1}&1.861&
\makebox[0pt][r]{0}\makebox[0pt][l]{.6}&0.00& 
\makebox[0pt][r]{$>0$}\makebox[0pt][l]{}&
\makebox[0pt][r]{$\approx$ 0}\makebox[0pt][l]{.3}&
Insulator&$-$\\ 
\makebox[0pt][r]{A}\makebox[0pt][l]{2}&1.863&
\makebox[0pt][r]{1}\makebox[0pt][l]{.6}&0.00&
\makebox[0pt][r]{$>0$}\makebox[0pt][l]{}&
\makebox[0pt][r]{1}\makebox[0pt][l]{}&
Insulator&$-$\\ 
\makebox[0pt][r]{A}\makebox[0pt][l]{3}&1.912&
\makebox[0pt][r]{7}\makebox[0pt][l]{.7}&0.04&
\makebox[0pt][r]{1}\makebox[0pt][l]{.7}&
\makebox[0pt][r]{4}\makebox[0pt][l]{.1}&
\makebox[0pt][r]{0}\makebox[0pt][l]{.1}&
\makebox[0pt][r]{7}\makebox[0pt][l]{.5}\\
\makebox[0pt][r]{A}\makebox[0pt][l]{4}&2.210&
\makebox[0pt][r]{15}\makebox[0pt][l]{.5}&0.15&
\makebox[0pt][r]{$-2$}\makebox[0pt][l]{.3}&
\makebox[0pt][r]{7}\makebox[0pt][l]{}&
\makebox[0pt][r]{5}\makebox[0pt][l]{.7}&
\makebox[0pt][r]{6}\makebox[0pt][l]{.4}\\
\makebox[0pt][r]{A}\makebox[0pt][l]{5}&2.232&
\makebox[0pt][r]{19}\makebox[0pt][l]{.1}&0.23&
\makebox[0pt][r]{$-2$}\makebox[0pt][l]{.4}&
\makebox[0pt][r]{8}\makebox[0pt][l]{}&
\makebox[0pt][r]{9}\makebox[0pt][l]{.6}&
\makebox[0pt][r]{5}\makebox[0pt][l]{.2}\\
\makebox[0pt][r]{B}\makebox[0pt][l]{1}&1.933&
\makebox[0pt][r]{7}\makebox[0pt][l]{.8}&0.04&
\makebox[0pt][r]{1}\makebox[0pt][l]{.4}&
\makebox[0pt][r]{4}\makebox[0pt][l]{}&
\makebox[0pt][r]{0}\makebox[0pt][l]{.2}&
\makebox[0pt][r]{7}\makebox[0pt][l]{.4}\\
\makebox[0pt][r]{B}\makebox[0pt][l]{2}&2.004&
\makebox[0pt][r]{11}\makebox[0pt][l]{.9}&0.09&
\makebox[0pt][r]{$-1$}\makebox[0pt][l]{.2}&
\makebox[0pt][r]{5}\makebox[0pt][l]{.5}&
\makebox[0pt][r]{2}\makebox[0pt][l]{.3}&
\makebox[0pt][r]{7}\makebox[0pt][l]{.8}\\
\makebox[0pt][r]{B}\makebox[0pt][l]{3}&2.076&
\makebox[0pt][r]{12}\makebox[0pt][l]{.0}&0.09&
\makebox[0pt][r]{$-1$}\makebox[0pt][l]{.3}&
\makebox[0pt][r]{6}\makebox[0pt][l]{}&
\makebox[0pt][r]{2}\makebox[0pt][l]{.6}&
\makebox[0pt][r]{7}\makebox[0pt][l]{.4}\\
\makebox[0pt][r]{B}\makebox[0pt][l]{4}&2.219&
\makebox[0pt][r]{18}\makebox[0pt][l]{.5}&0.22&
\makebox[0pt][r]{$-2$}\makebox[0pt][l]{.6}&
\makebox[0pt][r]{8}\makebox[0pt][l]{.0}&
\makebox[0pt][r]{8}\makebox[0pt][l]{.9}&
\makebox[0pt][r]{5}\makebox[0pt][l]{.5}\\
\makebox[0pt][r]{B}\makebox[0pt][l]{5}&2.290&
\makebox[0pt][r]{19}\makebox[0pt][l]{.8}&0.25&
\makebox[0pt][r]{$-2$}\makebox[0pt][l]{.7}&
\makebox[0pt][r]{8}\makebox[0pt][l]{}&
\makebox[0pt][r]{10}\makebox[0pt][l]{.4}&
\makebox[0pt][r]{4}\makebox[0pt][l]{.9}\\
\makebox[0pt][r]{B}\makebox[0pt][l]{6}&2.362&
\makebox[0pt][r]{19}\makebox[0pt][l]{.8}&0.25&
\makebox[0pt][r]{$-2$}\makebox[0pt][l]{.6}&
\makebox[0pt][r]{8}\makebox[0pt][l]{}&
\makebox[0pt][r]{10}\makebox[0pt][l]{.4}&
\makebox[0pt][r]{5}\makebox[0pt][l]{.0}\\
\makebox[0pt][r]{B}\makebox[0pt][l]{7}&2.434&
\makebox[0pt][r]{22}\makebox[0pt][l]{}&0.32&
\makebox[0pt][r]{$-2$}\makebox[0pt][l]{.3}&
\makebox[0pt][r]{9}\makebox[0pt][l]{}&
\makebox[0pt][r]{13}\makebox[0pt][l]{.8}&
\makebox[0pt][r]{3}\makebox[0pt][l]{.9}\\ 
\end{tabular}
\end{table} 
\begin{table}
\caption{Critical exponent $\mu$ and prefactor $\sigma^*$ 
for the metal-insulator transition.  
Values of $\sigma^*$ normalized to Mott's minimum metallic 
conductivity defined by Eq.~(\protect{\ref{eq:s_min}}) is also listed.}
\label{tab:pre}
\begin{tabular}{cccccc} 
 & & Magnetic induction &  & $\sigma^*$ & \\ 
System & Ref. & (T) & $\mu$ & (10$^2$ S/cm) & 
$\sigma^*/\sigma_{\rm min}$ \\ \hline 
$^{70}$Ge:Ga & \protect{\ref{Wat98}}&
\makebox[0pt][r]{0}\makebox[0pt][l]{}&
\makebox[0pt][r]{0.50}\makebox[0pt][l]{$\;\pm\;$0.04}&
\makebox[0pt][r]{0}\makebox[0pt][l]{.4}&
\makebox[0pt][r]{6}\makebox[0pt][l]{} \\
Ge:Sb & \protect{\ref{Oot88}}&
\makebox[0pt][r]{0}\makebox[0pt][l]{}&
\makebox[0pt][r]{$\approx0.9$}\makebox[0pt][l]{}&
\makebox[0pt][r]{0}\makebox[0pt][l]{.6}&
\makebox[0pt][r]{9}\makebox[0pt][l]{} \\
Si:B & \protect{\ref{Dai92}}&
\makebox[0pt][r]{0}\makebox[0pt][l]{}&
\makebox[0pt][r]{0.65}\makebox[0pt][l]{$\;^{+\;0.05}_{-\;0.14}$}&
\makebox[0pt][r]{1}\makebox[0pt][l]{.5}&
\makebox[0pt][r]{8}\makebox[0pt][l]{} \\
Si:P & \protect{\ref{Ros83}}&
\makebox[0pt][r]{0}\makebox[0pt][l]{}&
\makebox[0pt][r]{$\approx0.48$}\makebox[0pt][l]{$\;-\;$0.55}&
\makebox[0pt][r]{3}\makebox[0pt][l]{}&
\makebox[0pt][r]{$\approx13$}\makebox[0pt][l]{} \\
Si:P & \protect{\ref{Dai93}}&
\makebox[0pt][r]{0}\makebox[0pt][l]{}&
\makebox[0pt][r]{$0.58$}\makebox[0pt][l]{$\;\pm\;$0.08}&
\makebox[0pt][r]{3}\makebox[0pt][l]{}&
\makebox[0pt][r]{$14$}\makebox[0pt][l]{} \\
$^{70}$Ge:Ga & this work &
\makebox[0pt][r]{4}\makebox[0pt][l]{}&
\makebox[0pt][r]{1.1}\makebox[0pt][l]{$\;\pm\;$0.1}&
\makebox[0pt][r]{0}\makebox[0pt][l]{.6}&
\makebox[0pt][r]{8}\makebox[0pt][l]{} \\
$^{70}$Ge:Ga & this work &
\makebox[0pt][r]{5}\makebox[0pt][l]{}&
\makebox[0pt][r]{1.1}\makebox[0pt][l]{$\;\pm\;$0.1}& \makebox[0pt][r]{0}\makebox[0pt][l]{.6}&
\makebox[0pt][r]{8}\makebox[0pt][l]{} \\
$^{70}$Ge:Ga & this work &
\makebox[0pt][r]{6}\makebox[0pt][l]{}&
\makebox[0pt][r]{1.1}\makebox[0pt][l]{$\;\pm\;$0.1}& \makebox[0pt][r]{0}\makebox[0pt][l]{.6}&
\makebox[0pt][r]{9}\makebox[0pt][l]{} \\
$^{70}$Ge:Ga & this work &
\makebox[0pt][r]{7}\makebox[0pt][l]{}&
\makebox[0pt][r]{1.1}\makebox[0pt][l]{$\;\pm\;$0.1}& \makebox[0pt][r]{0}\makebox[0pt][l]{.7}&
\makebox[0pt][r]{10}\makebox[0pt][l]{} \\
$^{70}$Ge:Ga & this work &
\makebox[0pt][r]{8}\makebox[0pt][l]{}&
\makebox[0pt][r]{1.1}\makebox[0pt][l]{$\;\pm\;$0.1}& \makebox[0pt][r]{0}\makebox[0pt][l]{.7}&
\makebox[0pt][r]{10}\makebox[0pt][l]{} \\
Ge:Sb & \protect{\ref{Oot88}}&
\makebox[0pt][r]{4}\makebox[0pt][l]{}&
\makebox[0pt][r]{$\approx1.0$}\makebox[0pt][l]{}&
\makebox[0pt][r]{0}\makebox[0pt][l]{.6}&
\makebox[0pt][r]{8}\makebox[0pt][l]{} \\
Si:B & \protect{\ref{Dai92}}&
\makebox[0pt][r]{7}\makebox[0pt][l]{.5}&
\makebox[0pt][r]{1.0}\makebox[0pt][l]{$\;^{+\;0.10}_{-\;0.20}$}&
\makebox[0pt][r]{1}\makebox[0pt][l]{.7}&
\makebox[0pt][r]{9}\makebox[0pt][l]{} \\
Si:P & \protect{\ref{Dai93}}&
\makebox[0pt][r]{8}\makebox[0pt][l]{}&
\makebox[0pt][r]{$0.86$}\makebox[0pt][l]{$\;\pm\;$0.15}&
\makebox[0pt][r]{3}\makebox[0pt][l]{}&
\makebox[0pt][r]{$15$}\makebox[0pt][l]{} \\
\end{tabular}
\end{table} 
\begin{table}
\caption{Universality classes for the metal-insulator transition and 
values of the critical exponent $\nu$ for the correlation length 
according to Ref.~\protect{\ref{Bel94}}. 
The values are given for $d=2+\varepsilon$ dimensions except 
for class~G, where an approximate value for $d=3$ is given.}
\label{tab:UC}
\begin{tabular}{ccc} 
Symbol & Symmetry breaker & $\nu$ \\ \hline 
MF & Magnetic field & $[1+O(\varepsilon)]/\varepsilon$ \\
MI & Magnetic impurities & $[1+O(\varepsilon)]/\varepsilon$ \\
SO & Spin-orbit scattering & $[1+O(\varepsilon)]/\varepsilon$ \\
G & None & $\cong 0.75$ \\
\end{tabular}
\end{table} 

\begin{references}
\bibitem{Lee85}For a review, see P. A. Lee and T. V. Ramakrishnan, 
        Rev. Mod. Phys. {\bf 57}, 287 (1985). 
\bibitem{Bel94}For a review, see D. Belitz and T. R. Kirkpatrick, 
        Rev. Mod. Phys. {\bf 66}, 261 (1994). \label{Bel94} 
\bibitem{Ros83} T. F. Rosenbaum, R. F. Milligan, M. A. Paalanen, 
G. A. Thomas, R. N. Bhatt, and W. Lin, Phys. Rev. B {\bf 27}, 7509 (1983).
\label{Ros83}
\bibitem{Sha89}W. N. Shafarman, D. W. Koon, and T. G. Castner, Phys. Rev. 
B {\bf 40}, 1216 (1989). \label{Sha89}
\bibitem{Lon84}A. P. Long and M. Pepper, J. Phys. C {\bf 17}, 
        L425 (1984).
\bibitem{Ion91}A. N. Ionov, M. J. Lea, and R. Rentzsch, Pis'ma Zh. Eksp. 
Teor. Fiz. {\bf 54}, 470 (1991) [JETP Lett. {\bf 54}, 473 (1991)]. 
\bibitem{Ito96}K. M. Itoh, E. E. Haller, J. W. Beeman, W. L. Hansen, 
J. Emes, L. A. Reichertz, E. Kreysa, T. Shutt, A. Cummings, W. Stockwell, 
B. Sadoulet, J. Muto, J. W. Farmer, and V. I. Ozhogin, 
Phys. Rev. Lett {\bf 77}, 4058 (1996). \label{Ito96}
\bibitem{Wat98}M. Watanabe, Y. Ootuka, K. M. Itoh, and E. E. Haller, 
        Phys. Rev. B {\bf 58}, 9851 (1998).   \label{Wat98}
\bibitem{Mac81}A. MacKinnon and B. Kramer, Phys. Rev. Lett. {\bf 47}, 
1546 (1981). \label{Mac81}
\bibitem{Sle97} K. Slevin and T. Ohtsuki, Phys. Rev. Lett. {\bf 78},
 4085 (1997). \label{Sle97}
\bibitem{Cha86}J. Chayes, L. Chayes, D. S. Fisher, and T. Spencer, Phys. 
Rev. Lett. {\bf 57}, 2999 (1986). \label{Cha86}
\bibitem{Paz97}For a theoretical attempt to show that $\nu$ can be less 
than $2/3$ in three dimensions, see, for example, 
F. P\'{a}zm\'{a}ndi, R. T. Scalettar, and G. T. Zim\'{a}nyi, 
        Phys. Rev. Lett. {\bf 79}, 5130 (1997).  
\bibitem{Weg76}F. J. Wegner, Z. Phys. B {\bf 25}, 327 (1976); {\em ibid.}
        {\bf 35}, 207 (1979). \label{Weg76}
\bibitem{Kir93}T. R. Kirkpatrick and D. Belitz, Phys. Rev. Lett. 
        {\bf 70}, 974 (1993). \label{Kir93}
\bibitem{Oot88}Y. Ootuka, H. Matsuoka, and S. Kobayashi, in {\em 
Anderson Localization}, edited by T. Ando and H. Fukuyama 
(Springer-Verlag, Berlin, 1988), p.~40.  \label{Oot88}
\bibitem{Ros89}T. F. Rosenbaum, S. B. Field and R. N. Bhatt, 
        Europhys. Lett. {\bf 10}, 269 (1989).
\bibitem{Dai92}P. Dai, Y. Zhang, and M. P. Sarachik, Phys. Rev. B 
        {\bf 45}, 3984 (1992).  \label{Dai92}
\bibitem{Dai93}P. Dai, Y. Zhang, S. Bogdanovich, and 
M. P. Sarachik, Phys. Rev. B {\bf 48}, 4941 (1993).  \label{Dai93}
\bibitem{Bog97}S. Bogdanovich, P. Dai, M. P. Sarachik, V. Dobrosavljevic, 
        and G. Kotliar, Phys. Rev. B {\bf 55}, 4215 (1997).  
\bibitem{Hal84}E. E. Haller, N. P. Palaio, M. Rodder, W. L. Hansen, and 
        E. Kreysa, in {\em Proceedings of the 4th International 
        Conference on Neutron Transmutation 
        Doping of Semiconductor Materials}, edited by R. D. Larrabee 
       (Plenum, New York, 1984), pp.~21--36.  
\bibitem{Par88}I. S. Park and E. E. Haller, J. Appl. Phys. {\bf 64}, 
        6775 (1988).  
\bibitem{Hal90}E. E. Haller, Semicond. Sci. Technol. {\bf 5}, 319 
        (1990).  
\bibitem{Ito93}K. Itoh, W. L. Hansen, E. E. Haller, J. W. Farmer, and 
        V. I. Ozhogin, in {\em Proceedings of the 5th International 
        Conference on Shallow Impurities in Semiconductors}, 
        edited by T. Taguchi [Mater. Sci. Forum {\bf 117 \& 118}, 
        117 (1993)].
\bibitem{Shi88}See, for example, F. Shimura, {\em Semiconductor Silicon 
        Crystal Technology} (Academic Press, San Diego, 1988), 
        pp.~159--161.  
\bibitem{Kit86}C. Kittel, {\em Introduction to Solid State Physics}
        (John Wiley \& Sons, New York, 1986) 6th~ed., chap.~8.
\bibitem{Khm81}D. E. Khmel'nitskii and A. I. Larkin, Solid State 
        Commun. {\bf 39}, 1069 (1981).
\end{references}
\end{document}